\documentclass[prl,twocolumn,showpacs,superscriptaddress]{revtex4}
\usepackage{graphicx}
\usepackage{color}

\newcommand{\notes}[1]{}

\newcommand{\beq}{\begin{equation}}
\newcommand{\eeq}{\end{equation}}
\newcommand{\beqnn}{\begin{equation*}}
\newcommand{\eeqnn}{\end{equation*}}
\newcommand{\beqas}{\begin{eqnarray*}}
\newcommand{\eeqas}{\end{eqnarray*}}
\newcommand{\beqa}{\begin{eqnarray}}
\newcommand{\eeqa}{\end{eqnarray}}
\newcommand{\p}{\partial}

\begin{document}

\title{Onset of a Propagating Self-Sustained Spin Reversal Front in a Magnetic System}

\author{P. Subedi}
\affiliation{Department of Physics, New York University, New York, New York 10003, USA}
\author{S. V\'elez}
\affiliation{Grup de Magnetisme, Dept. de F\'isica, Universitat de Barcelona, Spain}
\author{F. Maci\`a}
\affiliation{Department of Physics, New York University, New York, New York 10003, USA}
\author{S. Li}
\affiliation{Department of Physics, City College of New York, CUNY, New York, New York
10031, USA}
\author{M. P. Sarachik}
\affiliation{Department of Physics, City College of New York, CUNY, New York, New York
10031, USA}
\author{\\ J. Tejada}
\affiliation{Grup de Magnetisme, Dept. de F\'isica, Universitat de Barcelona, Spain}
\author{S. Mukherjee}
\author{G. Christou}
\affiliation{Department of Chemistry, University of Florida, Gainesville, Florida 32611,
USA}
\author{A. D. Kent}
\affiliation{Department of Physics, New York University, New York, New York 10003, USA}

\begin{abstract}
The energy released in a magnetic material by reversing spins  as they relax toward equilibrium can lead to a dynamical instability that ignites self-sustained rapid relaxation along a deflagration front that propagates at a constant subsonic speed.  Using a trigger heat pulse and transverse and longitudinal magnetic fields, we investigate and control the crossover between thermally driven magnetic relaxation and magnetic deflagration in single crystals of Mn$_{12}$-acetate.
 \end{abstract}
\maketitle

Raising the temperature of a flammable substance can ignite combustion, an exothermic reaction between a substance and an oxidizer that results in a chemically modified substance \cite{Glassman}. Deflagration is a self-sustained combustion that propagates at subsonic speed via thermal conduction; locally burning substance increases the temperature of adjacent unburnt substance and ignites it.  Deflagration is a dynamic instability of a system governed by local reactions and diffusion. These systems are ubiquitous in nature--from cell growth to epidemics. The study of these nonlinear dynamical systems reveals rich phenomenology--including traveling waves, dissipative solitons \cite{Akhmediev}, and self-organized patterns \cite{HohenbergRMP,Moses98}. 

Magnetic relaxation is a diffusion-reaction process that can develop an instability and proceed as a magnetic deflagration. It can occur in a magnetic material  prepared in a metastable spin configuration; here the reaction is the reversal of spins that release Zeeman energy and the diffusion serves to transmit the energy to adjacent material.  The phenomenon has been observed as fast magnetization jumps in molecular magnets \cite{suzuki,quantumdeflagration}, manganites \cite{macia2007} and intermetallic compounds \cite{velez2010}. In these systems, spins that are in a metastable state at low temperature slowly relax to a lower energy state releasing heat. In some circumstances, the heat cannot be compensated by thermal diffusion, resulting in an instability that gives rise to a front of rapidly reversing spins traveling through the sample at constant speed.

In this Letter we study the instability that leads to magnetic deflagration in a thermally driven Mn$_{12}$-Ac crystal. We explore how to control the crossover between slow magnetic relaxation and rapid, self-sustained magnetic deflagration using magnetic fields. Based on earlier work \cite{maciaepl2006,Garanin2007}, we propose a simple model and perform numerical simulations that provide a good description of the experimental findings.

\begin{figure}[t]
\centering

\includegraphics[width=\linewidth]{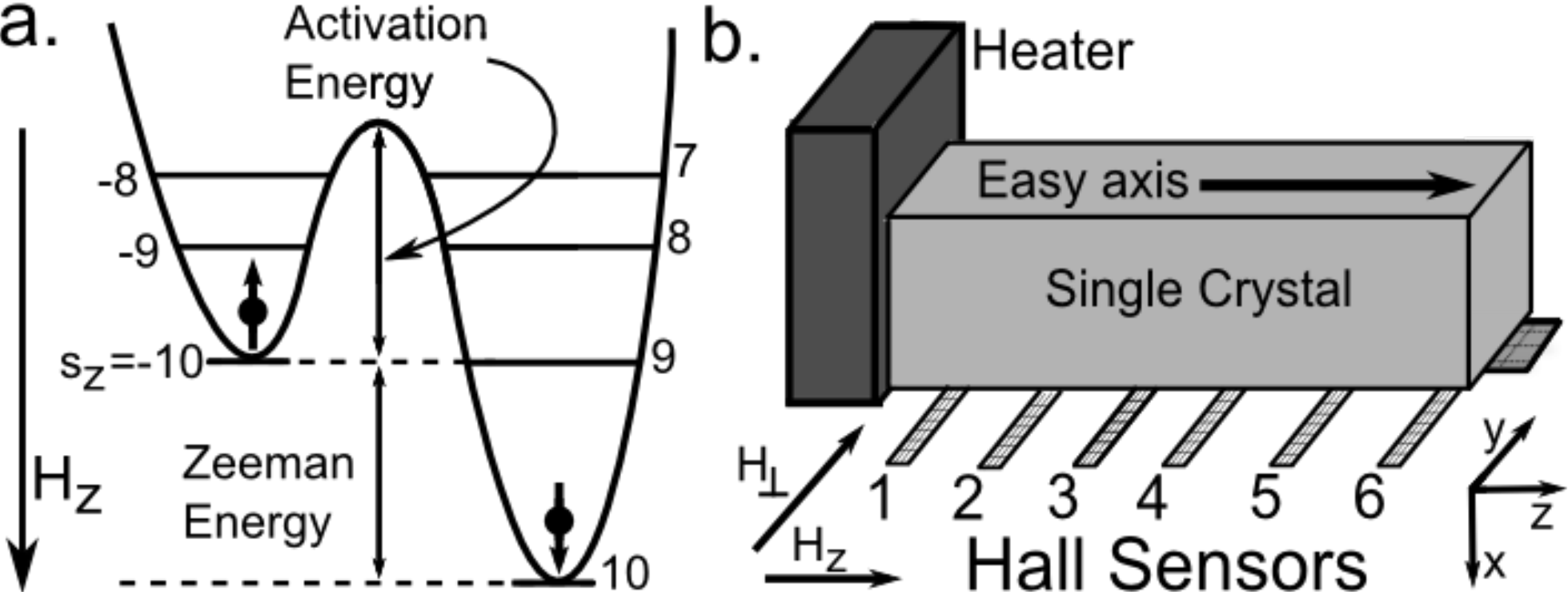}
\caption{a) The potential energy with discrete energy levels of  Mn$_{12}$-Ac spin Hamiltonian, Eq.\,\ref{hamintro4}. b) Schematic of the sample, the Hall sensors, the heater, and the directions of the applied magnetic fields.}
\label{fig1}
\end{figure}

Mn$_{12}$-Ac molecules are magnetically bistable \cite{sessoli1993} due to their high $S=10$ spin and a magnetic anisotropy that provides a large energy barrier between spin-up and spin-down states (see Fig.\,\ref{fig1}a). In the presence of an applied field $\mathbf{H}=(H_\perp,H_z)$ the simplest spin Hamiltonian  has the form:
\begin{equation}
\mathcal{H}= -DS_z^2-g\mu_B H_z S_z - g\mu_B H_\perp S_\perp. 
\label{hamintro4}
\end{equation}
The first term defines the anisotropy barrier ($DS^2 \approx 65$ K) and the second and third terms are the Zeeman energy corresponding to the field applied parallel and perpendicular to the anisotropy axis. As shown in Fig.\,\ref{fig1}a, the energy barrier between the metastable and the stable state is the activation energy, $U$, and the energy released by the reversing spins is the Zeeman energy, $\Delta E = 2g \mu_BH_zS_z$. The transverse field $H_\perp$ mixes eigenstates of $S_z$ reducing the activation energy and leading to relaxation by quantum tunneling of magnetization, particularly at specific {\it resonant} values of magnetic field ($H_z=kD/(g\mu_B) \simeq  0.45k$ T, for integer $k$), corresponding to the crossings of energy levels with opposite spin-projections. The longitudinal  field $H_z$ biases the potential well also reducing the activation energy and increasing the relaxation rate.

The abrupt complete reversal of the magnetic moment of Mn$_{12}$-Ac crystals was first reported by Paulsen and Park \cite{Paulsen} and later shown to take the form of a propagating spin reversal front \cite{suzuki}.  Magnetic deflagration has been studied by sweeping the magnetic field or by raising the crystal's temperature.  Thresholds for both temperature \cite{Mchugh2007} and applied Zeeman fields \cite{macia2009} were found to be governed by quantum laws due to quantum tunneling of magnetization \cite{JonathanPRL1996}. In the experiments reported below, the spin reversal was triggered using a heat pulse at one end of the sample.

\begin{figure}[tb]
\centering
\includegraphics[width=\linewidth]{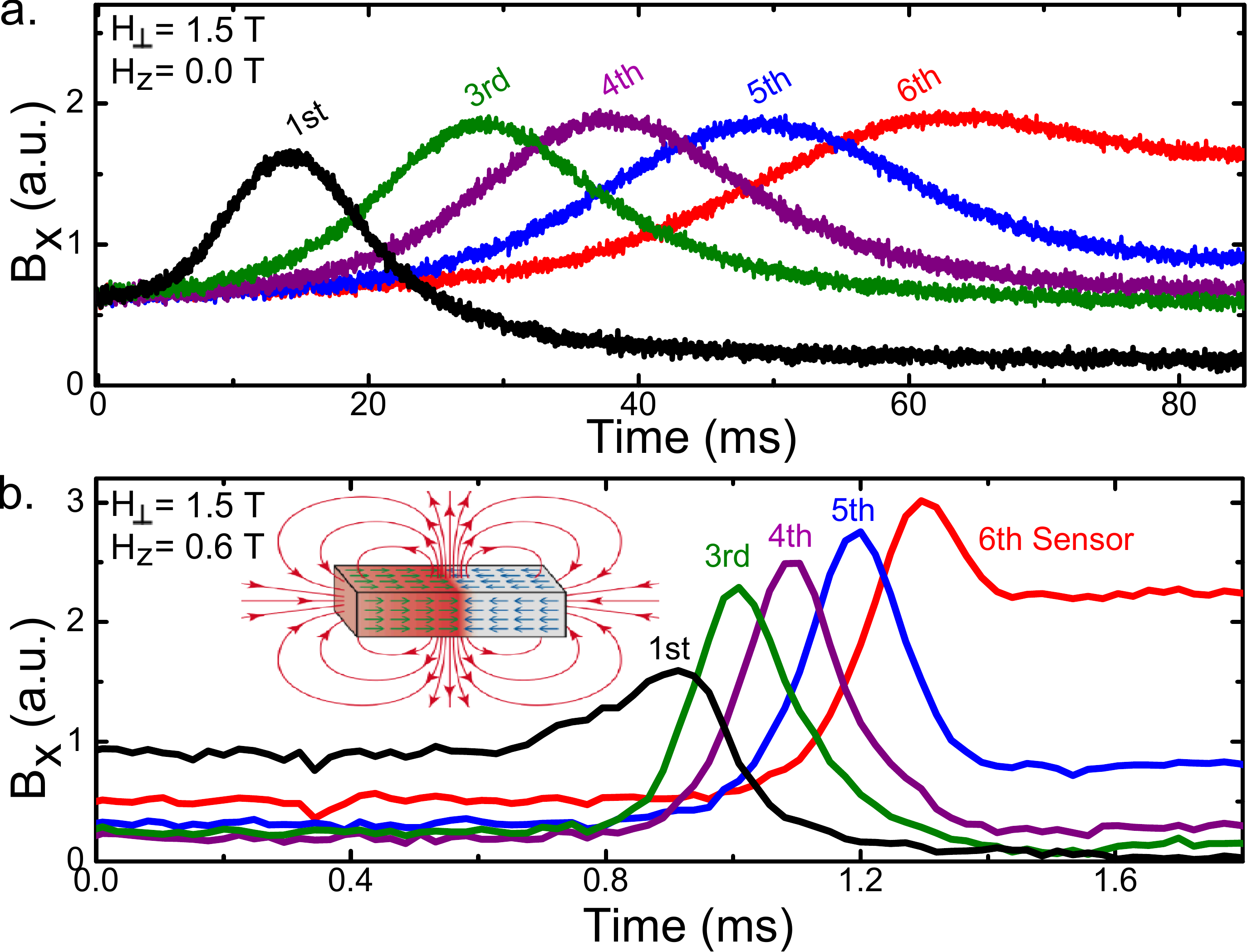}
\caption{Signals of the Hall sensor arrays as a function of time. a) At zero bias field, the peaks are very broad, the propagation time is long and the speed is not constant.  b) At $H_z = 0.61$ T, the peaks are sharp and spin reversal propagates at higher and constant speed. The inset shows stray fields from the sample when the reversal front is in the middle of the crystal. The maximum $B_x$ is at the position of the front.}
\label{fig2}
\end{figure}

Magnetic measurements were performed on three Mn$_{12}$-Ac single crystals of dimensions $\sim 0.3 \times 0.3 \times 2.1$ mm$^3$, $0.35 \times 0.35 \times 1.75$ mm$^3$ and $0.4 \times 0.4 \times 1.6$ mm$^3$ (samples A, B and C, respectively).  The samples were mounted on a one-dimensional array of Hall sensors (active area $20 \times 100\ \mu$m$^2$ with 200 $\mu$m separation) (see Fig.\ \ref{fig1}b). The three samples show essentially the same behavior; here we show data for sample C.  Care was taken to align the sample and the Hall array (placed in the $y$-$z$ plane) relative to each other and relative to the magnet axes, as shown in Fig.\,\ref{fig1}b. The Hall sensors detect the stray field $B_x$ (see inset of Fig.\,\ref{fig2}b), which is maximum at the reversal front. Measurements were taken at $0.4$ K in a $^3$He refrigerator in a 3D vector superconducting magnet capable of producing bipolar bias fields up to 1~T and bipolar transverse fields up to 5~T. A 6~V, 30 ms  pulse was applied to a thin film heater (R $\approx$ 1.32 k$\Omega$ at 0.4 K) placed at one end of the sample to trigger spin reversal. A dc current of 20 $\mu$A was supplied to the array of hall sensors and the signal from each sensor was amplified by a factor of 1000, filtered, and recorded by a data acquisition card.

At base temperature ($0.4$ K), the crystal was prepared in a fully magnetized state so that all the spins were aligned in one direction.  A fixed $H_\perp$ was applied and $H_z$ was swept to the desired field and held. At this low temperature the spins are essentially blocked by the strong magnetic anisotropy of Mn$_{12}$-Ac, and relax slowly toward equilibrium.  In this non-equilibrium state, a voltage pulse was then supplied to the heater to increase the temperature at one end of the crystal to initiate spin reversal. The same pulse was used to trigger the data acquisition card to record the magnetization signals of different Hall sensors during 1\,s. The procedure was repeated for several values of $H_z$ for a particular $H_\perp$.

Figure\,\ref{fig2} shows the time evolution of the Hall sensor signals when a heat pulse is applied at $t=0$ to sample C with the magnetization prepared as described above.  Two cases are considered: a zero bias field ($H_z = 0$ T, Fig.\,\ref{fig2}a), and a large bias field ($H_z = 0.61$ T, Fig.\,\ref{fig2}b) for the same $H_\perp=1.5$ T.  A propagating spin reversal region is observed in both cases; the change in magnetization always starts at the heater end and propagates towards the other end (from 1$^{st}$ to 6$^{th}$ sensor, Fig.\,\ref{fig1}b). As shown in Fig.\,\ref{fig2}, a peak in the signal corresponding to the enhancement of the local stray field ($B_x$) travels in the z-direction as the spin reversal propagates along the easy axis, ultimately reversing the magnetization of the entire crystal. The speed of propagation of the reversing spins can be determined from the time difference between pulses sensed by adjacent Hall sensors and their separation.


While the curves in Fig.\,\ref{fig2}a and Fig.\,\ref{fig2}b appear similar, closer inspection shows that the reversal process at low and high bias are distinctly different. At zero (or low) bias, the time interval between adjacent peaks is relatively long and increases as the spin-reversal propagates, indicating that the spin reversal is slowing down; the time to completion is on the order of $80$ ms; and the peaks are fairly broad and broaden further as the reversal proceeds.  Here the spin reversal propagates slowly and its speed decreases because the magnetization relaxes toward equilibrium following the heat pulse; its progress through the crystal is governed by thermal diffusivity.

By contrast, in the case of high bias shown in Fig.\,\ref{fig2}b, the time between adjacent peaks is constant and on the order of $100\ \mu$s, corresponding to a constant speed of $v \sim 2$ m/s, consistent with the data measured by Suzuki \emph{et al.} \cite{suzuki}; the total time for reversal of the full magnetization of the crystal is of the order of 1 ms; and the peaks in Fig.\,\ref{fig2}b are substantially sharper and do not broaden as they travel.  The constant speed in the case of a large $H_z$ is a signature of magnetic deflagration, a self-sustained process driven by the Zeeman energy released by the spins as they reverse.

\begin{figure}[t]
\centering
\includegraphics[width=\linewidth]{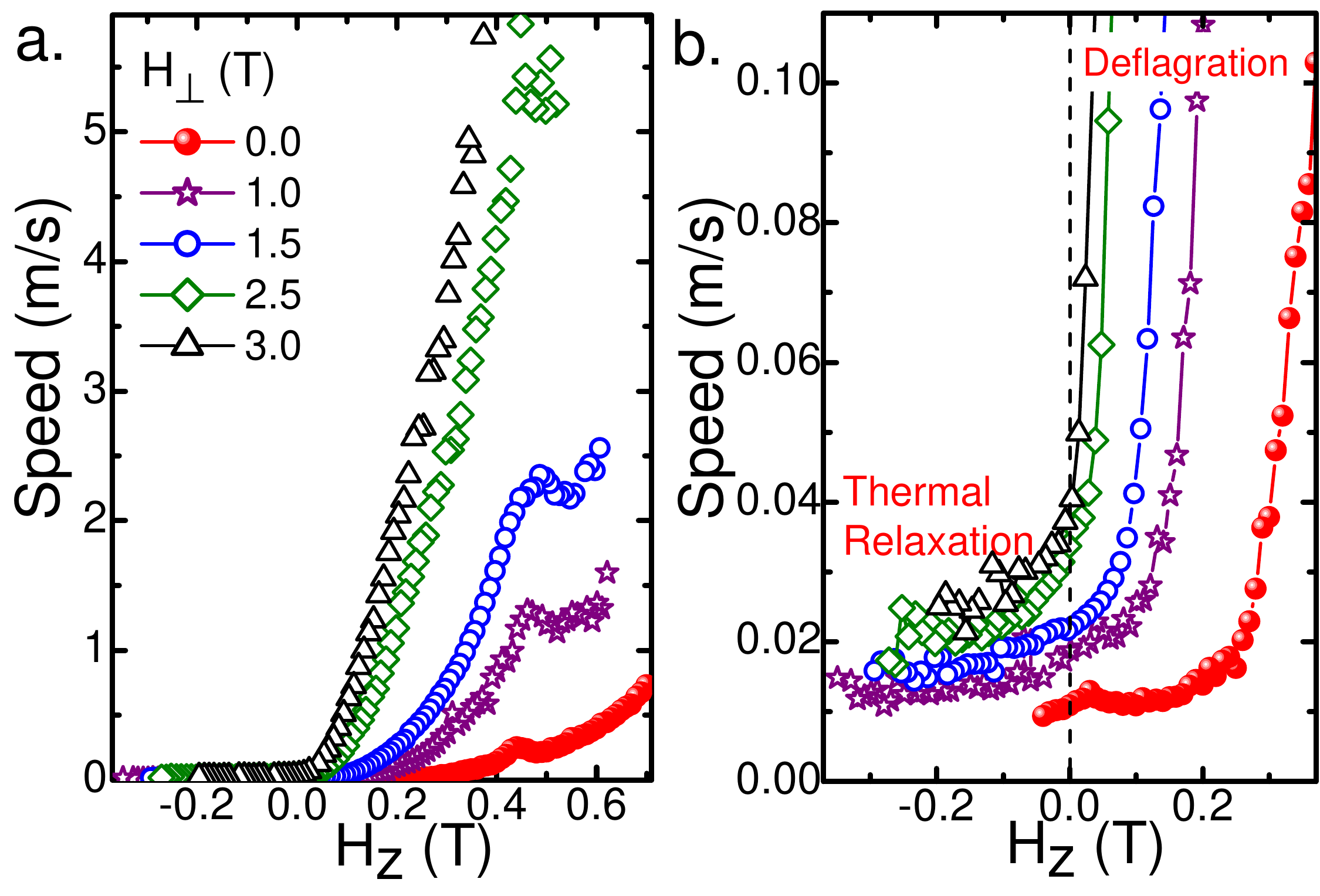}
\caption {a) The speed of propagation of the spin reversal as a function of $H_z$ for different $H_\perp$.  b) A blow-up of the data for small $H_z$. The speed is small and nearly constant at very small $H_z$ and increases abruptly when a deflagration develops.}
\label{fig3}
\end{figure}

We will now show that the crossover between these two regimes is surprisingly sharp.  The speed of propagation of the spin reversal fronts was determined from the time interval between maxima of the Hall sensors in the middle of the crystal (sensor 3 and 4) and their spatial separation. Note that while the same speed is obtained using any pair of Hall sensors for large $H_z$, the speed  varies from point to point at low $H_z$, and therefore between different Hall sensor pairs. As shown in Fig.\,\ref{fig3}, the speed of the spin reversal front changes abruptly at $H_z$ that depends on the transverse field.  For a given $H_\perp$, below a crossover bias field $H_{co}$ (for e.g., $H_{co}$ = 0.25 T for $H_\perp$ = 0 T), the speed of propagation of the reversing spins is nearly independent of the bias field. However, above the crossover, the speed increases suddenly and depends strongly on the bias field.  Figure\,\ref{fig3}b, a magnification of Fig.\,\ref{fig3}a where the vertical scale has been expanded by a factor of $100$, demonstrates this even more clearly.

The peaks in the speed at $H_z\sim0.45$ T shown in Fig.\,\ref{fig3}a are a clear manifestation of the important influence of quantum mechanics on the dynamics of the system.  These maxima are due to quantum tunneling at the resonant fields \cite{JonathanPRL1996,Hernandez1996} whenever spin states on opposite sides of the anisotropy barrier have nearly the same energy. As noted earlier, tunneling enhances relaxation and effectively reduces the anisotropy barrier. As we increase $H_\perp$ we increase the tunnel splitting between levels, which results in a broadening of the resonance steps.  A small peak can also be seen in Fig.\,\ref{fig3}b at the zero bias resonance. Note that the zero-field resonance is slightly shifted ($\sim 40$\ mT) due to internal dipolar fields \cite{McHugh2009,BoPRB2010}.

To better understand the crossover between the two regimes described above, we develop a simple model based on previous theory \cite{maciaepl2006,Garanin2007}. The diffusion-reaction system that describes the time evolution of the magnetization, $m$, towards equilibrium, $m_{eq}$, entails two processes: the Zeeman energy (i.e. heat) released into the sample by the spins as they reverse and the heat diffusion through the sample. The dynamical system of nonlinear partial differential equations reads:
\begin{equation}
\dot{m}=-\Gamma(m-m_{\text{eq}}) \quad\quad
\dot{T}=\dot{m}\ \Delta E/C+\nabla\cdot\kappa\nabla T.
\label{DS}
\end{equation}
Here $T$ is the temperature, $\kappa$ is the thermal diffusivity, $C$ is the heat capacity and $\Gamma=\Gamma_0\exp\left[-U(h_\perp,h_z)/(k_BT)\right]$ is the relaxation rate. For fields that are much smaller than the anisotropy field ($h_i=g\mu_B H_i/2DS \ll 1$) \cite{Garanin1997}:
\beq
U(h_\perp,h_z)\approx DS^2 (1-h_z)^2\left[1-2h_\perp\frac{(1-h_z^2)^{1/2}}{(1-h_z)^2}\right].
\label{barrier}
\eeq

For some applied fields and initial conditions, the thermal diffusivity cannot compensate for the increase in temperature due to the Zeeman energy released by the reversing spins; the sample temperature then rapidly increases and a magnetic deflagration develops. To study the ignition of magnetic deflagration, we consider a system of spins blocked in a given metastable magnetic state, increase its temperature (e.g., from 0.5 to 5\,K), and study the resulting nucleation process. Let us consider here for simplicity the temperature evolution in a nucleation volume, independent of coordinates,
\beq
\dot{T}=\dot{m}\ \Delta E/C-2\kappa/R^2(T-T_{0})
\label{insteq}
\eeq
where $T$ is the temperature of the volume under study, $T_{0}$ is the temperature of the rest of the sample and $2R$ is the characteristic size of the nucleation volume. The negative term in Eq.\ \ref{insteq} is linear with temperature while the reaction term increases exponentially with temperature ($\dot{m} \propto \exp\left[-U/(k_BT)\right]$); there is thus a competition between these two terms and stationary solutions ($\dot{T}=0$) correspond to temperatures when the dissipation equals the released energy.

\begin{figure}[t]
\centering
\includegraphics[width=\linewidth]{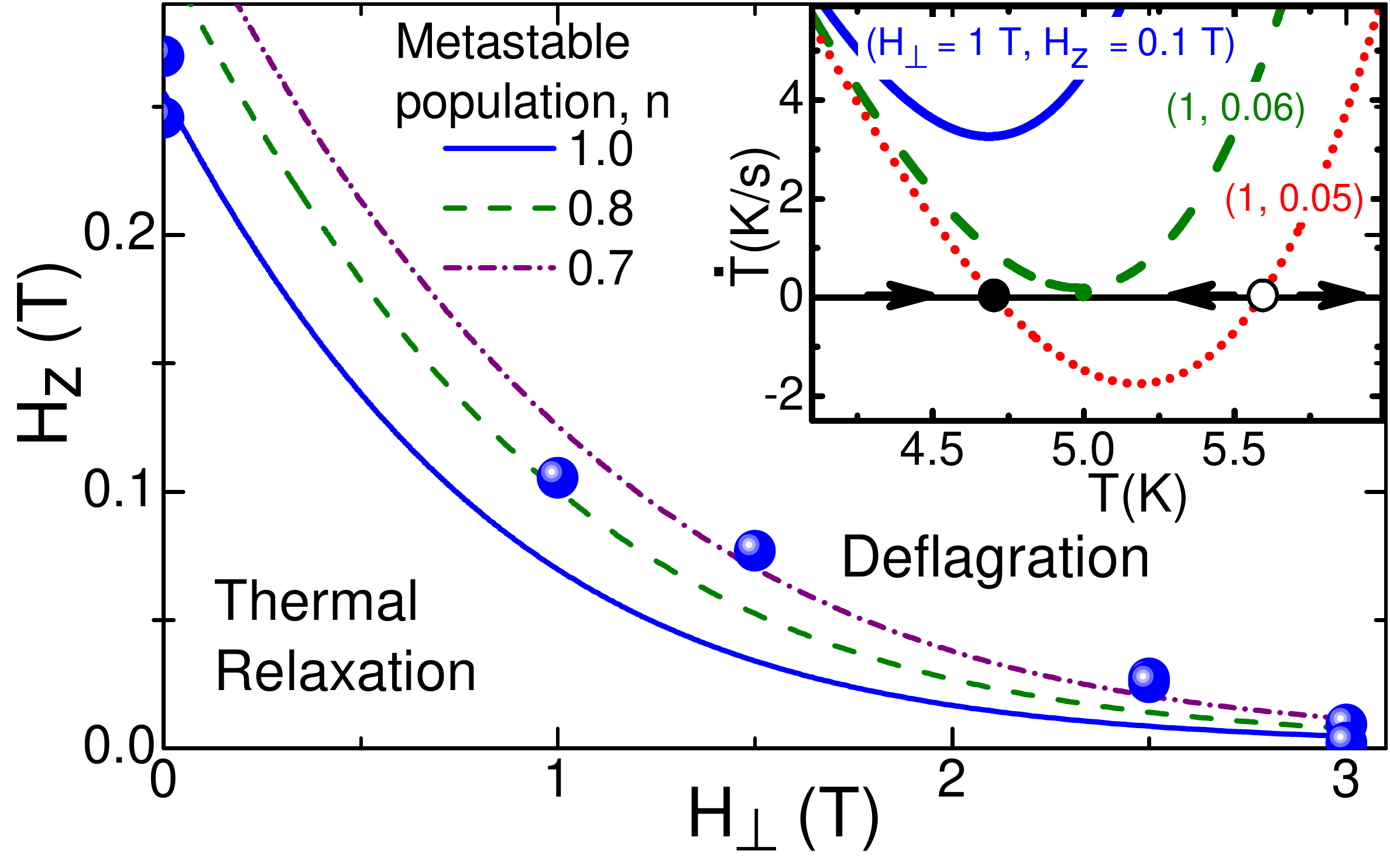}
\caption{Main panel: The boundary curves separating thermal and Zeeman regimes for various $n$. The blue dots are $H_{co}$, derived from the data in Fig.\,\ref{fig3}.  The inset plots $\dot{T}$ as a function of $T$ for three different $\mathbf{H}$. The solid and open black circles indicate stable and unstable solutions for the $(1, 0.05)$ T field case.}
\label{fig4}
\end{figure}

As shown in the inset of Fig.\,\ref{fig4}, some magnetic field configurations ($\mathbf{H}=(H_\perp,H_z)$) corresponding to small bias fields have stationary solutions (red dotted curve). As we vary $\mathbf{H}$ (i.e., we vary the relaxation rate, $\dot{m}$), the stationary points move closer to each other and eventually merge (green dashed curve) and disappear. When there is no stationary solution (blue solid line), the temperature derivative is always positive leading to an instability--a magnetic deflagration may develop. The condition for the crystal to lose stability at given $T_{0}$ and $\mathbf{H}$ is $\dot{T}>0$ and its crossover points are found by solving
\beq
\dot{T}=0,\quad\quad \p\dot{T}/\p T= 0.
\label{instCond}
\eeq

At a given $H_\perp$, for $H_z < H_{co}$, the energy released during the spin reversing process is small (or even zero in the case of zero bias). The dynamics of the propagating front observed in this case (Fig.\ \ref{fig2}a) are determined by the diffusion of the heat supplied by the heater pulse; this corresponds to a thermal regime. Here the reaction term in Eq.\ \ref{insteq} is negligible and the equation has a stable solution at a temperature close to the sample temperature. Neglecting the reaction term in the full set of Eqs.\,\ref{DS}, it is clear that the heat diffuses and the magnetization locally follows the sample temperature.

For $H_z\geq H_{co}$, the energy released by the reversing spins is large, the reaction term dominates over the diffusion term and Eq.\,\ref{insteq} displays an instability. In this Zeeman regime, the released energy drives the spin relaxation to a deflagration with steady propagation, where the speed of the propagating front of reversing spins strongly depends on both $H_z$ and $H_\perp$.

The sharp transition in propagation speed from thermal to Zeeman regimes as a function of the applied magnetic field is given by the condition in Eqs.\,\ref{instCond}. $H_\perp$ varies the activation barrier and does not change the released energy; it affects only $\dot{m}$ in the reaction term. $H_z$ affects both $\dot{m}$ and $\Delta E$.

In the main panel of Fig.\,\ref{fig4}, we plot the boundaries that separate the thermal regime from the Zeeman regime derived from Eq.\,\ref{instCond} together with the experimental points (blue dots) obtained from Fig.\,\ref{fig3}. The different curves denote different initial metastable spin population ($n=(M_s-m)/(2M_s)$, where $M_s$ is saturation magnetization) that determines the initial state \footnote{In an applied transverse field when the zero resonance is crossed, some of the spins at metastable state relax to the stable state and the initial population (fuel) decreases.}. For fields above the boundary curve, the reaction term dominates and the crystal develops a magnetic deflagration; below the curves, thermal diffusion dominates and the spins evolve toward equilibrium through slow thermal relaxation. For the calculations we have bounded $2R$ by the smallest sample dimension of 0.4\ mm, the heat capacity was taken to be $C/k_{B}=1$ (i.e. $C = 8.3$ J/(mol K)), close to a measured value \cite{GomesC}, and the thermal diffusivity was estimated to be $\kappa \approx 1\times 10^{-4} \ m^2/s$ \footnote{The thermal diffusivity value was estimated from the thermal relaxation measurements performed in the same sample.}, which agrees well with previously used values \cite{jmhapl2006}. The free parameter is the temperature, $T_0$, to which the sample is heated by the voltage pulse; the best fit is obtained for $T_0=5$\ K.

The smoothness and reproducibility of the data in Fig.\,\ref{fig3} suggest that the appearance of instabilities leading to a magnetic deflagration when raising the temperature of the Mn$_{12}$-Ac crystal is a well controlled process. Once the crystal is in the Zeeman regime the speed of the magnetic deflagration can be controlled by the strength of the applied field. Additionally, equal propagation speeds are obtained for different combinations of applied fields, as seen in Fig.\,\ref{fig3} where a particularly speed can be seen to correspond to different magnetic field configurations. Larger bias fields would produce larger amounts of heat released and higher temperatures for the overall process; while, large transverse fields that speed up the process by the same amount would heat the sample much less.

We have measured the dynamics of spin reversal fronts in Mn$_{12}$-Ac crystals in the presence of both bias and transverse magnetic fields. We have shown how these applied fields can be used to control an instability that separates slow magnetic relaxation from rapid, self-sustained magnetic deflagration.

We have also expanded the range of conditions under which magnetic deflagration has been observed, particularly to the case of small bias fields, which corresponds to small energy release in the deflagration process. This is a particularly interesting limit in which instabilities in the front are predicted to occur, such as pattern formation and oscillations of the front position as it propagates \cite{modestovprb}. In the presence of large transverse fields, where quantum tunneling dominates over thermal effects, the instabilities in the magnetization dynamics may result in supersonic fronts \cite{Garanin2012} and lead to a deflagration-to-detonation transition \cite{modestovprl,Decelle}. The present experiments thus  provide opportunities to study--in a nondestructive manner--a large variety of instabilities in magnetic systems.

We acknowledge illuminating discussions with Dov Levine. ADK and PS acknowledges support by NSF-DMR-1006575 and NYU. FM acknowledges support from a Marie Curie IOF 253214. MPS acknowledges support from NSF-DMR-0451605. Support for GC was provided under grant CHE-0910472. SV acknowledges financial support from the Ministerio de Educaci\'on, Cultura y Deporte de Espa\~na and JT acknowledges financial support from ICREA Academia.

\bibliographystyle{apsrev}
\bibliography{deflagration}

\end{document}